# GALAXIES AT $z = 2$: EXTENSIONS AROUND RADIO-QUIET QSOs

ITZIAR ARETXAGA, B.J. BOYLE, R.J. TERLEVICH

*Royal Greenwich Observatory, Madingley Road, Cambridge CB3 0EZ, U.K.*

**Abstract**

We have been conducting an imaging survey to detect host galaxies of radio-quiet QSOs at high redshift ($z \approx 2$), in order to compare them with those of radio-loud objects. Six QSOs were observed in the $R$ passband with the auxiliary port of the 4.2m WHT of the *Observatorio de Roque de los Muchachos* in August 1994. The objects were selected to be bright ($M_B \lesssim -28$ mag) and have bright stars in the field, which could enable us to define the point spread function (PSF) accurately. The excellent seeing of La Palma ($< 0.9$ arcsec thoughout the run) allowed us to detect extensions to the nuclear PSFs around three (one radio-loud and two radio-quiet) QSOs, out of 4 suitable targets. The extensions are most likely due to the host galaxies of these QSOs, with luminosities of at least $3 - 7\%$ of the QSO luminosity. The most likely values for the luminosity of the host galaxies lie in the range $6 - 18\%$ of the QSO luminosity. Our observations show that, if the extensions we have detected are indeed galaxies, extraordinary massive and luminous galaxies are not only a characteristic of radio-loud objects, but of QSOs as an entire class.

## 1. Analysis of the observations

For each QSO field, we defined a 2-dimensional PSF using the brightest of the closest stellar companion to the QSO. We then scaled the PSF to match the luminosity of the QSO and other nearby stars over the same region, and subtracted the scaled PSF from them. The remaining residuals in the non-PSF stars provided an accurate check of the validity of the subtraction process. We accepted the PSF subtraction if the residuals in the non-PSF stars accounted for less than $1\sigma$ of the Poisson noise expected from the subtraction technique. We have detected residuals in excess of $3\sigma$ in the following QSOs: 1630.5+3749 ($4\sigma$), PKS 2134+008 ($3\sigma$) and Q 2244−0105 ($3.7\sigma$). All the residuals show a 'doughnut' shape with a well of negative counts in the centre. This indicates that there is a flatter component below the PSF in the centres of the QSOs, from which the nuclear (PSF) contribution has been over-subtracted. The left-hand panel in Fig. 1 shows the fields of one of the radio-quiet QSOs (1630.5+3749) around which we have detected an extension. The right-hand panel of Fig. 1 shows the same field after subtracting the corresponding PSF from both QSO and stars.

To estimate the true luminosity of these systems, we have subtracted smaller amounts of the PSF in order either a) to produce zero counts in the center of the residuals or b) to achieve a flat-top profile with no depression in the center. We regard these quantities as a lower limit (3–7% of the QSO luminosity) and a best estimate (6–18% of the QSO luminosity), respectively, of the total luminosity of these extended components. In all cases, the FWHM of the flat-top residual profiles are significantly larger than the FWHM of the stars in each field: $1''.05$ vs $0''.7$ for 1630.5+3749, $0''.8$ vs. $0''.7$ for PKS 2134+008 and $0''.84$ vs. $0''.7$ for Q 2244−0105.

## 2. Discussion

Out of a sample of one radio-loud and three radio-quiet QSOs with suitable PSF stars, we have been able to detect extensions in three cases [1]. The best estimates for luminosities of these systems lie between 6–18% of the total QSO luminosity. Host galaxies of radio-loud QSOs at $z \sim 2$ which comprise up to 20% of the QSO $R$ band luminosity have already been detected [3] but, to date, no other detection has been made of hosts of high-redshift radio-quiet QSOs.

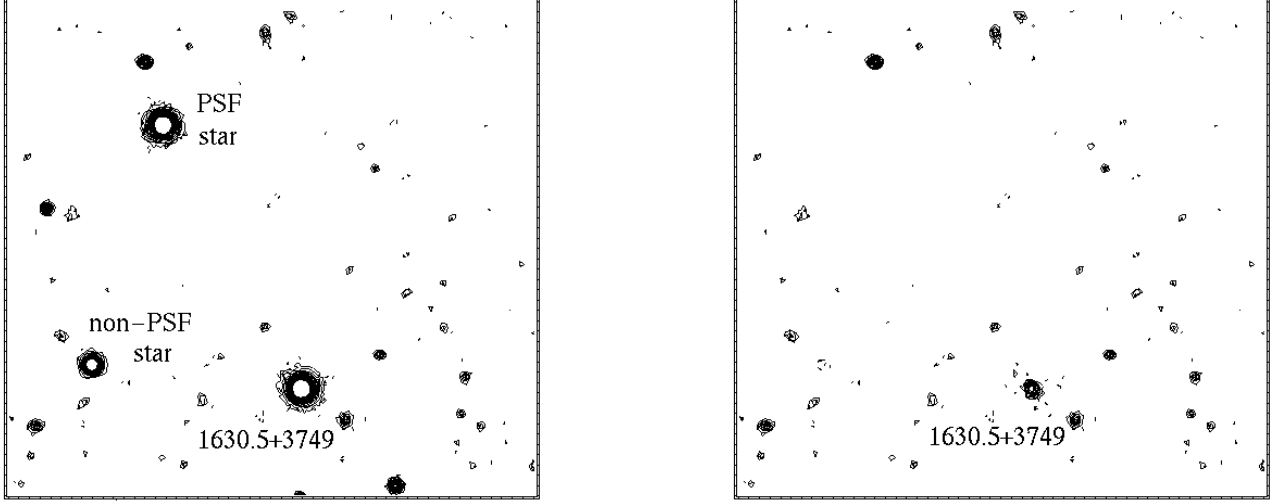

Figure 1: $50'' \times 50''$ field of the QSOs 1630.5+3749, before and after subtracting the PSF.

For radio-loud QSOs, the red colours of the extensions [3] favour a host galaxy origin of the light, rather than scattering by dust or electrons in the halo of the QSOs. Although we lack colour information, the similarility between the nuclear to extended component luminosity ratio in radio-loud QSOs with that observed here, leads us to believe that the extensions we are detecting are also due to light from the host galaxy. Indeed, scattering models require the presence of a powerful transverse radio-jet [2] which is unlikely to be present in either the radio-quiet QSOs, or the core-dominated radio-loud QSO around which we have detected extensions.

Galaxies as luminous as the extensions detected here have already been found in the imaging survey of radio-loud QSOs carried out by Lehnert et al. [3]. Four of the objects of their sample, with similar redshift to those in our sample, show 'fuzz' around the PSF of the nucleus. In the observed $R$ frame the absolute magnitude of this 'fuzz' ranges from $-25.6$ to $-26.9$ mag ($H_0 = 50$, $q_0 = 0.5$), as derived from the $B$ and $K$ colours they report. Their galaxies are $\sim 12$ times more luminous than present-day giant ellipticals and $\sim 2.5 - 3$ mag brighter than the hosts of low redshift QSOs. Our observations show that, if the extensions we have detected are indeed galaxies, massive and luminous galaxies are not only a characteristic of radio-loud objects, but of QSOs as an entire class.

Indeed, the radio-loud QSO studied in this sample does not exhibit a significantly larger or a more luminous extension than those of radio-quiet QSOs. One of the radio-quiet QSOs exhibits no significant evidence for any extension. This is in contrast to the work of Lowenthal et al. [4] who have established upper limits to the observed $K$ luminosity of the host galaxies of a sample of high redshift ($z \approx 2.5$) radio-quiet QSOs to be $\sim 2$ mag less luminous than those of radio-loud QSOs [3].

**Acknowledgments:** IA's work is supported by the EEC HCM fellowship ERBCHBICT941023